\documentclass[10pt,a4paper]{article}
\usepackage{amsmath}
\usepackage{amsmath,latexsym,amssymb,amsfonts}
\usepackage[dvips]{graphicx}
\usepackage{bm}
\usepackage{xcolor}

\begin{document}

\title{\textbf{Acausality in Nonlocal Gravity Theory}}

\author{\textsc{Ying-li Zhang$^{a,b}$}\footnote{{\tt yingli{}@{}bao.ac.cn}},
 \textsc{Kazuya Koyama$^{b}$}\footnote{{\tt kazuya.koyama{}@{}port.ac.uk}},
 \textsc{Misao Sasaki$^{c}$}\footnote{{\tt misao{}@{}yukawa.kyoto-u.ac.jp}}\;\;
and \textsc{Gong-Bo Zhao$^{a,b}$}\footnote{{\tt gbzhao{}@{}nao.cas.cn}}\\
\textit{\small{$^{a}$National Astronomy Observatories, Chinese Academy of Science,}}\\
\textit{\small{Beijing 100012, People's Republic of China}}\\
\textit{\small{$^{b}$Institute of Cosmology and Gravitation,
 University of Portsmouth, Portsmouth PO1 3FX, UK}}\\
\textit{\small{$^{c}$Yukawa Institute for Theoretical Physics,
 Kyoto University, Kyoto 606-8502, Japan}}
}

\maketitle

\begin{abstract}

We investigate the nonlocal gravity theory by deriving nonlocal equations of motion using the traditional variation principle in a homogeneous background. We focus on a class of models with a linear nonlocal modification term in the action. It is found that the resulting equations of motion contain the advanced Green's function, implying that there is an acausality problem. As a consequence, a divergence arises in the solutions due to contributions from the future infinity unless the Universe will go back to the radiation dominated era or become the Minkowski spacetime in the future. We also discuss the relation between the original nonlocal equations and its biscalar-tensor representation and identify the auxiliary fields with the corresponding original nonlocal terms. Finally, we show that the acusality problem cannot be avoided by any function of nonlocal terms in the action.
\end{abstract}

\begin{flushright}
{\tt YITP-16-3}
\end{flushright}

\newpage

\tableofcontents

\newpage

\section{Introduction}

The nonlocal gravity theory was initially proposed as a ``filter'' to eliminate the contribution of the cosmological constant to the spacetime curvature so that it might provide a possible way to relieve the cosmological constant problem~\cite{ArkaniHamed:2002fu,Nojiri:2010pw,ZS:2011,BNOS:2011}. As far as cosmological studies are concerned, a model with nonlocal modifications was proposed by Deser and Woodard in 2007~\cite{Deser:2007jk}. At the level of action, the nonlocal correction term takes the form of $Rf(\Box^{-1}R)$, in which the dimensionless combination $\Box^{-1}R$ is tiny during the radiation-dominated era but gradually increases in the matter-dominated epoch. Hence, this theory could help relieve the ``fine-tuning problem'' of the dark energy without introducing any small mass scale. Based on this model, the cosmological correspondences were studied extensively (e.g. see Refs.~\cite{Joukovskaya:2007nq}--\cite{DFKMP:2015}). However, it was found that although at the background level, the evolution of the nonlocal gravity could be designed to be indistinguishable from that of $\Lambda$CDM model~\cite{DW:2009}, studies of the structure formation would disfavor this model~\cite{PD:2013,PD2:2013}.

Nevertheless, this negative result does not totally rule out the possibility of including nonlocal corrections in the action. Recently, there appear a series of studies of nonlocal modifications: in Refs.~\cite{JMM:2013,M:2013,FMM:2014}, a term proportional to $g_{\mu\nu}\Box^{-1}R$ was introduced into the field equations. It was found that in this model, a mass term could be introduced without any reference metric~\cite{JMM:2013,MT:2013}. Moreover, its equation of state (EoS) is less than $-1$, hence this model could mimic the phantom dark energy~\cite{M:2013,FMM:2014}, while studies of its linear perturbations showed that this model was statistically comparable with $\Lambda$CDM model~\cite{NT:2014}. Another model was proposed by introducing a term proportional to $R\Box^{-2}R$ into the action~\cite{MM:2014}, with its cosmological perturbations studied in~\cite{DFKKM:2014} which also gave a positive result. Besides these two interesting models, there are also discussions on other related topics, e.g. interpretations of dark matter as nonlocal effects from the General Relativity (GR)~\cite{SW:2003,Bar:2011,Bar:2012}.

On the other hand, several theoretical aspects of a theory with nonlocal terms remain to be clarified. For instance, in order to transform the original
integro-differential equation into differential equations, the nonlocal gravity theory is often written into a biscalar-tensor theory by introducing a scalar field $\psi\equiv\Box^{-1}R$ with a Langrangian multiplier. In this case, the number of degrees of freedom in this theory becomes ambiguous. It was found that in the corresponding biscalar-tensor theory, there would appear a ``ghost-like'' mode so that the theory could become unhealthy~\cite{Nojiri:2010pw,ZS:2011,BNOS:2011,deS:2015}. However, it was argued in~\cite{M:2013,FMM:2014gh} that in the biscalar-tensor theory, the Green's function for $\psi$ should be defined in the way where the initial conditions remove the homogeneous solution which satisfies $\Box\psi_\mathrm{hom}=0$. In this sense, when $\psi$ is quantized, the creation and annihilation coefficients vanish so that $\psi$ is not a ``free field'', hence the ``ghost-like'' mode is physically irrelevant.

Another problem is the appearance of acausality in this theory. As discussed in~\cite{Deser:2007jk,Woodard:2014,SW:2003,FMM:2014gh}, under the replacement $x\leftrightarrow x'$, the retarded Green's function $G_R(x', x)$ becomes an advanced one $G_A(x, x')$. Hence, in the Minkowskian background, for a class of theories which contain nonlocal operators acting on scalar fields, it is expected that the advanced Green's function cannot be eliminated in the equations of motion (EOM) obtained by the traditional variation principle. One of the consequences is that the future information is needed in order to find the solutions, which may imply acusality problems of the theories.

In this paper, we consider the acausality problem arising from the nonlocal gravity theory. A similar problem may appear in a class of modified gravity theories that contain nonlocal operators. We start from a linear nonlocal gravity action and derive the EOM in its original formulation by the variation principle. We find that the variation principle will symmetrize the property of Green's function in the EOM, i.e., no matter whether the nonlocal operator is defined by the retarded Green's function or the advanced one in the action, both of them symmetrically appear in the EOM. This means that the advanced Green's function cannot be eliminated by any construction of functions of the nonlocal operator in the action. Hence, future information is needed to find the solutions, i.e. the acusality problem appears in the nonlocal gravity theory. This could imply that the nonlocal gravity theory is not well-defined, or it is not a fundamental theory to derive the causal nonlocal equations.

In most literature, especially for the numerical analysis, the analysis is done in the biscalar-tensor representation. We make a comparison of the original EOM to its biscalar-tensor representation and identify the extra scalars with the non-local terms in the original formulation. We find that one of the additional scalars would be identified to the term associated with advanced Green's function. In the biscalar-tensor representation, when we solve the EOM for the extra scalars, we reverse the relationship $\psi=\Box^{-1}R$ to the second-order differential equation $\Box\psi=R$. Hence, in the solution for this second-order differential equation, there appears a homogeneous solution that makes the solution different from the original one, i.e. $\psi_{\rm sol}=\Box^{-1}R+U_{\rm hom}$, where $U_{\rm hom}$ is the homogeneous solution satisfying $\Box U_{\rm hom}=0$, which causes the ``ghost'' instability problem~\cite{Nojiri:2010pw,ZS:2011,BNOS:2011,deS:2015}.

This paper is organised as follows. In Section~\ref{sec2}, we start our study with a simple example where the nonlocal operator acts on a scalar field in the action. From this simple study, it is straightforward to show the appearance of the advanced Green's function in the corresponding equation of motion. In Section~\ref{EOMoriginal}, we focus on a class of nonlocal gravity models where the nonlocal modification term is linear in the action. In Section 3.1, by using the traditional variation principle, we derive the nonlocal equations of motion in the homogeneous background and find that both the retarded and advanced Green's function will appear symmetrically. In Section 3.2, we use the trace equation to demonstrate that the advanced Green's function cannot be eliminated from the EOM. In Section 3.3, we consider the relationship between the original nonlocal equation and its biscalar-tensor representation. We identify the additional scalars with their original nonlocal terms. Especially, in the homogeneous background, we demonstrate
that the difference between two representations is caused by a homogeneous solution and discuss its consequence. In Section~\ref{general}, we generalise the analysis to a model involving a general function of both advanced and retarded Green's functions in its action. We find that the acusality problem cannot be avoided by constructing any function of the non-local operator in the action. In the Appendix, as a supplement to Section 2, we show the appearance of the advanced Green's function in the EOM assuming the FLRW metric without using the property that $G_R(x, x')\leftrightarrow G_A(x', x)$ under the replacement of variables $x\leftrightarrow x'$.

\section{A simple example: scalar field with nonlocal operator}\label{sec2}
Before studying the nonlocal gravity theory, it would be useful to consider the case where the nonlocal operator acts on a scalar field $\phi(x)$ in the action. Let us consider the simplest example where the nonlocal operator appears linearly in the action:
\begin{align}\label{scalaraction}
S_{\phi}&=\int d^4x\sqrt{-g(x)}\phi(x)\left(\Box^{-1}\phi\right)[x]\,.
\end{align}
where the nonlocal operator $\Box^{-1}$ operating on an arbitrary function $f(x)$ is defined by the integration of Green's function $G(x,
x')$ as follows
\begin{align}
\left(\Box^{-1}f\right)[x]\equiv\int d^4x'\sqrt{-g(x')}f(x')G(x,
x')\,.
\end{align}

An important property of the Green's function is that, under the replacement of variables $x\leftrightarrow x'$, the retarded Green's function changes to the advanced one $G_R(x, x')\leftrightarrow G_A(x', x)$, and vice versa. Then the problem arises that when one varies the nonlocal term in (\ref{scalaraction}), the advanced Green's function appears when the action is defined by a retarded one~\cite{Deser:2007jk,Woodard:2014,SW:2003,FMM:2014gh}:
\begin{align}
&~~~\int d^4x\sqrt{-g(x)}\phi(x)\left(\Box^{-1}_R\frac{\delta
\phi}{\delta
\phi(y)}\right)[x]\nonumber\\
&=\int\int d^4x~d^4x'\sqrt{-g(x)}\sqrt{-g(x')}\phi(x)G_R(x,
x')\frac{\delta \phi(x')}{\delta \phi(y)}\nonumber\\
&=\int\int d^4x'~d^4x\sqrt{-g(x')}\sqrt{-g(x)}\phi(x')G_R(x',
x)\frac{\delta \phi(x)}{\delta \phi(y)}\nonumber\\
&=\int d^4x\sqrt{-g(x)}\delta(x-y)\int
d^4x'\sqrt{-g(x')}\phi(x')G_R(x', x)\nonumber\\
&=\int d^4x\sqrt{-g(x)}\delta(x-y)\int
d^4x'\sqrt{-g(x')}\phi(x')G_A(x, x')\nonumber\\
&=\int d^4x\sqrt{-g(x)}\delta(x-y)\left(\Box^{-1}_A\phi\right)[x]\,. \label{scalareg}
\end{align}

Hence, variation of the nonlocal action (\ref{scalaraction}) with respect to the scalar field $\phi(x)$ symmetries the property of Green's functions, regardless whether the nonlocal operator in the action is defined by the retarded or advanced Green's function:
\begin{align}\label{scalareq}
\frac{\delta S_\phi}{\delta\phi(y)}=\sqrt{-g(y)}\left(\Box^{-1}_R\phi+\Box^{-1}_A\phi\right)[y]\,.
\end{align}
This implies that the advanced Green's function cannot be eliminated in the EOM for the scalar field. In the Appendix, the same result is obtained when the background spacetime is described by the FLRW metric without using the property of the Green's function.

Correspondingly, one may expect the similar result in a nonlocal gravity theory described by the action containing a non-local term
\begin{align}
S\supset S_{NL}=\int d^4x\sqrt{-g(x)}R(x)\left(\Box^{-1}R\right)[x]\,.
\end{align}
by casting $\phi(x)=R(x)$. The situation would seem to be different since the nonlocal operator $\Box^{-1}$ contains the determinant $\sqrt{-g}$. Also when the variation principle is applied with respect to $g_{\mu\nu}$, one should vary the nonlocal operator itself from which extra terms will appear in the EOM. However, we show in the next section that a direct variation of a nonlocal gravity action will again symmetrise the retarded and advanced Green's function in the EOM.

\section{Linear Nonlocal gravity in a homogeneous geometry}\label{EOMoriginal}
As a simple example of the nonlocal gravity theory proposed in Ref.~\cite{Deser:2007jk}, in this section, we consider a specific nonlocal gravity theory where the nonlocal modification is linear in the action, i.e.
\begin{align}\label{actiong}
S_{NL}=\int d^4x\sqrt{-g(x)}R(x)\left(\Box^{-1}_RR\right)[x]\,.
\end{align}
where for definiteness, we define the nonlocal operator in the action by the retarded Green's function, denoted as $\Box^{-1}_R$.

\subsection{Original equations of motion}\label{EOMdeduce}
In this subsection, in the original frame (\ref{actiong}), we treat the metric $g_{\mu\nu}$ as the only variable and derive the EOM directly by the traditional variation principle. The variation can be expressed in the following way
\begin{align}
\frac{\delta S_{NL}}{\delta g^{\mu\nu}(\tilde{x})}=\Delta G_{\mu\nu}^{(1)}+\Delta G_{\mu\nu}^{(2)}+\Delta G_{\mu\nu}^{(3)}\,,
\end{align}
where
\begin{align}
\Delta G_{\mu\nu}^{(1)}&\equiv\int d^4x\frac{\delta\sqrt{-g(x)}}{\delta g^{\mu\nu}(\tilde{x})}R(x)\left(\Box^{-1}_RR\right)[x]\,,\\
\Delta G_{\mu\nu}^{(2)}&\equiv\int d^4x\sqrt{-g(x)}\frac{\delta R(x)}{\delta g^{\mu\nu}(\tilde{x})}\left(\Box^{-1}_RR\right)[x]\,,\\
\Delta G_{\mu\nu}^{(3)}&\equiv\int d^4x\sqrt{-g(x)}R(x)\frac{\delta\left(\Box^{-1}_RR\right)[x]}{\delta g^{\mu\nu}(\tilde{x})}\,.
\end{align}
We note that the variations in $\Delta G_{\mu\nu}^{(1)}$ and $\Delta G_{\mu\nu}^{(2)}$ are analogue terms in GR, so they can be calculated straightforwardly as
\begin{align}
\Delta G_{\mu\nu}^{(1)}&=-\frac{1}{2}\int d^4x\sqrt{-g(x)}~\delta^4(x-\tilde{x})~g_{\mu\nu}(x)R(x)\left(\Box^{-1}_RR\right)[x]\nonumber\\
&=-\frac{1}{2}\sqrt{-g}~g_{\mu\nu}R\left(\Box^{-1}_RR\right)\,,\label{part1res}\\
\Delta G_{\mu\nu}^{(2)}&=\int d^4x\sqrt{-g(x)}~\delta^4(x-\tilde{x})\bigg\{R_{\mu\nu}(x)\left(\Box^{-1}_RR\right)[x]
-\nabla_{\mu}\nabla_{\nu}\bigg(\left(\Box^{-1}_RR\right)[x]\bigg)+g_{\mu\nu}(x)R(x)\bigg\}\nonumber\\
&=\sqrt{-g}\bigg[R_{\mu\nu}\left(\Box^{-1}_RR\right)
-\nabla_{\mu}\nabla_{\nu}\left(\Box^{-1}_RR\right)+g_{\mu\nu}R\bigg]\,.\label{part2res}
\end{align}

However, $\Delta G_{\mu\nu}^{(3)}$ contains two pieces in which the property of Green's function will be changed:
\begin{align}
\Delta G_{\mu\nu}^{(3)}=\Delta G_{\mu\nu}^{(3-\mathrm{I})}+\Delta G_{\mu\nu}^{(3-\mathrm{II})}\,,
\end{align}
where
\begin{align}
\Delta G_{\mu\nu}^{(3-\mathrm{I})}&\equiv\int d^4x\sqrt{-g(x)}\int d^4x_1~\delta^4(x_1-\tilde{x})R(x)\frac{\delta\left(\Box^{-1}_RR\right)[x]}{\delta R(x_1)}\frac{\delta R(x_1)}{\delta g^{\mu\nu}(\tilde{x})}\nonumber\\
&=\int d^4x\sqrt{-g(x)}\bigg(R_{\mu\nu}(\tilde{x})
-\widetilde{\nabla}_{\mu}\widetilde{\nabla}_{\nu}+g_{\mu\nu}(\tilde{x})\widetilde{\Box}\bigg)\left(R(x)\frac{\delta\left(\Box^{-1}_RR\right)[x]}{\delta R(\tilde{x})}\right)\,,\\
\Delta G_{\mu\nu}^{(3-\mathrm{II})}&\equiv\int d^4x\sqrt{-g(x)}R(x)\left[\left(\frac{\delta\Box^{-1}_R}{\delta g^{\mu\nu}(\tilde{x})}\right)R\right][x]\,.\label{G3ii}
\end{align}
In fact, $\Delta G_{\mu\nu}^{(3-\mathrm{I})}$ is analogous to the scalar field case considered in Sec.~\ref{sec2} where the property of the Green's function changes. Hence, this term can be expressed as
\begin{align}
\Delta G_{\mu\nu}^{(3-\mathrm{I})}&= \bigg(R_{\mu\nu}(\tilde{x})
-\widetilde{\nabla}_{\mu}\widetilde{\nabla}_{\nu}+g_{\mu\nu}(\tilde{x})\widetilde{\Box}\bigg)\int d^4x\sqrt{-g(x)}\delta^4(x-\tilde{x})\left(\Box^{-1}_AR\right)[x]\nonumber\\
&=\sqrt{-g}\bigg[R_{\mu\nu}\left(\Box^{-1}_AR\right)
-\nabla_{\mu}\nabla_{\nu}\left(\Box^{-1}_AR\right)+g_{\mu\nu}R\bigg]\,,\label{part31res}
\end{align}
which is a symmetric counterpart of $\Delta G_{\mu\nu}^{(2)}$ shown in Eq.~(\ref{part2res}). Moreover, $\Delta G_{\mu\nu}^{(3-\mathrm{II})}$ is an extra term arising from the variation of the nonlocal operator itself. Generally speaking, it is difficult to analyse this term. In the following, we discuss it by assuming a homogeneous geometry, i.e. the Ricci scalar is only dependent on time: $R=R(t)$. Under this assumption, the nonlocal term can be expressed explicitly in the following way~\cite{Nonperturbative}
\begin{align}
\label{boxg}
\left(\Box^{-1}R\right)[t]&=-\int_{t_1}^t\frac{dt'}{\sqrt{-g(t')}}\int_{t_0}^{t'}dt''R(t'')\sqrt{-g(t'')}\,,
\end{align}
where $t_0, t_1\rightarrow-\infty$ corresponds to the retarded Green's function and $t_0, t_1\rightarrow+\infty$ gives the advanced Green's function. Taking into account that $\delta\sqrt{-g}=-\frac{1}{2}\sqrt{-g}g_{\mu\nu}\delta g^{\mu\nu}$, the variation of the retarded Green's function can be expressed as
\begin{align}\label{boxvar}
\left[\left(\frac{\delta\Box^{-1}}{\delta g^{\mu\nu}(\tilde{x})}\right)R\right][x]=&-\frac{1}{2}\int_{t_1}^tdt'\frac{g_{\mu\nu}(t')}{\sqrt{-g(t')}}\delta(t'-\tilde{t})\int_{t_0}^{t'}dt''\sqrt{-g(t'')}R(t'')\nonumber\\
&+\frac{1}{2}\int_{t_1}^t\frac{dt'}{\sqrt{-g(t')}}\int_{t_0}^{t'}dt''\sqrt{-g(t'')}g_{\mu\nu}(t'')R(t'')\delta(t''-\tilde{t})\,.
\end{align}
Thus, $\Delta G_{\mu\nu}^{(3-\mathrm{II})}$ can be calculated as
\begin{align}
\Delta G_{\mu\nu}^{(3-\mathrm{II})}&=\int d^4x\sqrt{-g(x)}R(x)\left[\left(\frac{\delta\Box^{-1}_R}{\delta g^{\mu\nu}(\tilde{x})}\right)R\right][x]\nonumber\\
&=-\frac{g_{\mu\nu}(\tilde{t})}{2\sqrt{-g(\tilde{t})}}\int_{\tilde{t}}^{+\infty}dt\sqrt{-g(t)}R(t)\int_{-\infty}^{\tilde{t}}dt''\sqrt{-g(t'')}R(t'')
-\frac{1}{2}\sqrt{-g}~g_{\mu\nu}R\left(\Box^{-1}_AR\right)\,,\label{part32res}
\end{align}
where we used a similar method with that to derive Eq.~(\ref{stepin2}) in Appendix to find the second term proportional to $\left(\Box^{-1}_AR\right)$, which comes from the second line of Eq.~(\ref{boxvar}). We note that this term is a symmetric counter term of $\Delta G_{\mu\nu}^{(1)}$ shown in Eq.~(\ref{part1res}). Hence, combining Eqs.~(\ref{part1res}), (\ref{part2res}), (\ref{part31res}) and (\ref{part32res}) together, we obtain all the terms arising from the variation of the action (\ref{actiong}) as
\begin{align}\label{eomorigin}
\frac{\delta S_{NL}}{\delta g^{\mu\nu}}&=\sqrt{-g}\left(R_{\mu\nu}-\frac{1}{2}g_{\mu\nu}R
-\nabla_{\mu}\nabla_{\nu}\right)\left[\left(\Box^{-1}_R+\Box^{-1}_A\right)R\right]+2\sqrt{-g}~g_{\mu\nu}R\nonumber\\
&-\frac{g_{\mu\nu}}{2\sqrt{-g}}\int_t^{+\infty}dt'\sqrt{-g(t')}R(t')\int_{-\infty}^tdt''\sqrt{-g(t'')}R(t'')\,.
\end{align}

It is obvious from this equation that for a simple class of models (\ref{actiong}) which are linear in the nonlocal operator in their actions, the corresponding equation of motion symmetrises the retarded and advanced Green's functions. In the original nonlocal theory, it seems difficult to understand the implication of the second line of Eq.~(\ref{eomorigin}) which originates from the variation of nonlocal operator itself (shown in Eq.~(\ref{part32res})). This term is understood as the cross product of the first order derivative of $\Box^{-1}_RR$ and $\Box^{-1}_AR$ in a biscalar-tensor presentation of the nonlocal theory as we will show in Section.~\ref{biscalartensor}.

\subsection{A simple case: the trace equation}
In order to understand whether Eq.~(\ref{eomorigin}) requires information to the infinite future or not, it would be useful to consider the trace equation. Taking a trace of the equation of motion, we can reduce Eq.~(\ref{eomorigin}) to the form:
\begin{align}\label{eomtrace}
g_{\mu\nu}\frac{\delta S_{NL}}{\delta g^{\mu\nu}}&=-\sqrt{-g}R\left[\left(\Box^{-1}_R+\Box^{-1}_A\right)R\right]+6\sqrt{-g}R\nonumber\\
&~~~-\frac{2}{\sqrt{-g}}\int_t^{+\infty}dt'\sqrt{-g(t')}R(t')\int_{-\infty}^tdt''\sqrt{-g(t'')}R(t'')\,.
\end{align}

We hope that the second line could be merged to the first line and eliminate the advanced Green's function. To see if this is possible, we should first notice that
\begin{align}\label{st1}
\sqrt{-g}R\Box^{-1}_RR&=\left[\partial_t\int_T^tdt'\sqrt{-g(t')}R(t')\right]\Box^{-1}_RR\nonumber\\
&=-\left[\int_T^tdt'\sqrt{-g(t')}R(t')\right]\partial_t(\Box^{-1}_RR)
+\partial_t\left[\Box^{-1}_RR\int_T^tdt'\sqrt{-g(t')}R(t')\right]\nonumber\\
&=\frac{1}{\sqrt{-g}}\int_T^tdt'\sqrt{-g(t')}R(t')\int_{-\infty}^tdt''\sqrt{-g(t'')}R(t'')
+\partial_t\left[\Box^{-1}_RR\int_T^tdt'\sqrt{-g(t')}R(t')\right]\,,
\end{align}
where $\partial_t\equiv d/dt$ and we used Eq.~(\ref{boxg}) in the last step. So it follows that
\begin{align}\label{str2}
-\sqrt{-g}R\left[\left(\Box^{-1}_R+\Box^{-1}_A\right)R\right]&=
-\frac{1}{\sqrt{-g}}\left[\int_T^tdt'\sqrt{-g}R\int_{-\infty}^tdt''\sqrt{-g}R+\int_{\tilde{T}}^tdt'\sqrt{-g}R\int_{+\infty}^tdt''\sqrt{-g}R\right]\nonumber\\
&~~~-\partial_t\left[\Box^{-1}_RR\int_T^tdt'\sqrt{-g(t')}R(t')+\Box^{-1}_AR\int_{\tilde{T}}^tdt'\sqrt{-g(t')}R(t')\right]\,,
\end{align}
where $T$ and $\tilde{T}$ are two integration boundaries. For simplicity, here we let $T\rightarrow+\infty$ and $\tilde{T}\rightarrow-\infty$ so that the first line of Eq.~(\ref{str2}) cancels with the second line of Eq.~(\ref{eomtrace}).~\footnote{Here we note that if one takes $T\rightarrow-\infty$ and $\tilde{T}\rightarrow+\infty$, there appears an extra contant term $\left(\int_{-\infty}^{+\infty}dt'\sqrt{-g}R\right)^2$, which gives a strong constraint that the Ricci scalar $R$ should vanish quickly when $t$ goes to infinite.} Hence, we obtain the following simpler expression:
\begin{align}
g_{\mu\nu}\frac{\delta S_{NL}}{\delta g^{\mu\nu}}=6\sqrt{-g}R-\partial_t\left[\Box^{-1}_RR\int_{+\infty}^tdt'\sqrt{-g(t')}R(t')+\Box^{-1}_AR\int_{-\infty}^tdt'\sqrt{-g(t')}R(t')\right]\,.
\end{align}
As can be observed above, the information for the future evolution is always required for the integrations in the EOM.

\subsection{The biscalar-tensor representation}\label{biscalartensor}
In this simple class of non-local gravity, we may
rewrite the action into a local form by introducing
a scalar field $\psi=\Box^{-1}_RR$ and a Langrangian multiplier $\xi$ as follows~\cite{Nojiri:2010pw,ZS:2011,BNOS:2011,Nojiri:2007nl1}
\begin{align}\label{bisca}
S&=\frac{1}{2\kappa^2}\int d^4 x \sqrt{-g}
 R\left(1 + \Box^{-1}_RR\right)\nonumber\\
&=\frac{1}{2\kappa^2}\int d^4 x \sqrt{-g}
\left[R\left(1 + \psi\right) -
\xi\left(\Box\psi - R\right)
\right]
 \nonumber\\
&=\frac{1}{2\kappa^2} \int d^4x \sqrt{-g}\left[R\left(1 +
\psi+\xi\right) + g^{\mu\nu}\partial_\mu \xi \partial_\nu \psi
\right] \,.
\end{align}
By varying the action with respect to $g_{\mu\nu}$, $\xi$ and
$\psi$, respectively, we obtain the field equations as
\begin{align}
0=&~\frac{1}{2}g_{\mu\nu} \left[R\left(1+
\psi+ \xi\right)
 + g^{\alpha\beta}\partial_\alpha \xi \partial_\beta \psi
\right]-R_{\mu\nu}\left(1+\psi+\xi\right) \nonumber\\
& -\frac{1}{2}\left(\partial_\mu \xi \partial_\nu \psi +
\partial_\mu \psi \partial_\nu \xi\right)
 -\left(g_{\mu\nu}\Box - \nabla_\mu \nabla_\nu\right)\left(\psi
 +\xi\right)\,,\label{st1}\\
0=&~R-\Box\psi\,,\label{st2}\\
0=&~R-\Box\xi\,.\label{st3}
\end{align}

Let us clarify the (in-)equivalence between the original nonlocal action (\ref{actiong}) and its biscalar-tensor presentation (\ref{bisca}). A question is how many degrees of freedom are there in the biscalar-tensor representation. Actually, in the biscalar-tensor presentation, one has already reversed the definition $\psi=\Box^{-1}_RR$ to obtain Eq.~(\ref{st2}). The difference between two representations thus appears in the degrees of freedom since the solution given by Eq.~(\ref{st2}) generally contains a homogeneous solution $U_{\rm hom}$ defined by $\Box U_{\rm hom}=0$, so that
\begin{align}\label{homfun}
\psi_{\rm sol}=\Box^{-1}R+U_{\rm hom}\,,
\end{align}
where $\psi_{\rm sol}$ is the solution obtained from the localized equation~(\ref{st2}). As observed in Ref.~\cite{MM:2014}, the homogeneous solution is uniquely fixed once we have specified the integration boundaries of $\Box^{-1}R$. This necessarily means that $U_{\rm hom}$ will never be a free field which could be expanded into plane waves. This can be seen clearly in the background described by the FLRW metric with a power-law solution $H(t)=s/t$ where $s=constant$. In this case, it immediately follows that
\begin{align}\label{Adxi}
\left(\Box^{-1}R\right)[t]&=-\int_{t_*}^t\frac{dt'}{\sqrt{-g(t')}}\int_{t_*}^{t'}dt''\sqrt{-g(t'')}R(t'')\nonumber\\
&=\frac{6s(2s-1)}{3s-1}\left\{\frac{1}{1-3s}\left[\left(\frac{t}{t_*}\right)^{1-3s}-1\right]-\ln\left(\frac{t}{t_*}\right)\right\}\,,
\end{align}
where the integration boundary $t_*$ corresponds to the boundary for the integration of the Green's function. On the other hand, in the localized form, the solution for $\psi$ can be directly obtained from the second-order differential equation~(\ref{st2}):
\begin{align}\label{psisol}
\psi_{\rm sol}(t)=\psi_1\left(\frac{t}{T_0}\right)^{1-3s}-\frac{6s(2s-1)}{3s-1}\ln\left(\frac{t}{T_0}\right)\,,
\end{align}
where $\psi_1$ and $T_0$ are two integration constants. Inserting Eqs.~(\ref{Adxi}) and (\ref{psisol}) into (\ref{homfun}),  the homogeneous solution $U_{\rm hom}$ can be explicitly expressed as
\begin{align}\label{homfunresult}
U_{\rm hom}&=\psi_{\rm sol}-\Box^{-1}_RR\nonumber\\
&=\psi_1\left(\frac{t}{T_0}\right)^{1-3s}+\frac{6s(2s-1)}{(3s-1)^2}\left(\frac{t}{t_*}\right)^{1-3s}+\beta\,,
\end{align}
where for simplicity, the constant $\beta$ is defined as
\begin{align}\label{betadef}
\beta\equiv\frac{6s(2s-1)}{3s-1}\left[\ln\left(\frac{t_*}{T_0}\right)+\frac{1}{1-3s}\right]\,.
\end{align}
From Eq.~(\ref{homfunresult}), it is straightforward to check that $\Box U_{\rm hom}=-\ddot U_{\rm hom}-3H\dot U_{\rm hom}=0$. Hence, the homogeneous solution $U_{\rm hom}$ will be fixed once the integration boundary $t_*$ is specified in this case.

%It should be note that the constant $t_R$ in Eq.~(\ref{homfunresult}) originates from the integration boundary for the retarded Green's function (\ref{Adxi}). This implies the reason why the acusality problem does not seem to appear in the biscalar-tensor presentation: the solution obtained from the localized second-order differential equation (\ref{st2}) already contains the homogeneous function which eliminates the integration boundary of $\Box^{-1}_RR$. As we have shown in Eq.~(\ref{homfun}), the correct way to recover the original definition is to remove the contribution from the homogeneous function by adding the counter term $\alpha U_{\rm hom}$. Thus, the spurious degree of freedom caused by the homogeneous function is absent in this treatment. Instead, one would immediately realize that the acausality problem is inevitable in the nonlocal gravity theory.

Practically, the biscalar-tensor presentation is useful especially when one is doing the numerical analysis since the original equation containing the differential-integral term (\ref{eomorigin}) is reduced to a set of second-order differential equations (\ref{st1})--(\ref{st3}). However,
to make sure that the biscalar-tensor representation is equivalent to the original nonlocal theory, the auxiliary fields $\psi$ and $\xi$ need to be identified with terms in the original nonlocal equation of motions.

For instance, in our linear nonlocal gravity case, comparing the localized equations (\ref{st1})--(\ref{st3}) to the original one (\ref{eomorigin}), one finds that the Langrangian multiplier $\xi$ plays the role of an advanced Green's function
\begin{align}\label{xipower}
\xi=\Box^{-1}_AR\,.
\end{align}
After identifying $\psi=\Box^{-1}_RR$ and $\xi=\Box^{-1}_AR$, it is useful to consider the cross terms between $\psi$ and $\xi$ in Eq.~(\ref{st1})
\begin{align}
I_{\mu\nu}\equiv\frac{\sqrt{-g}}{2}\bigg(\partial_\mu \xi \partial_\nu \psi +
\partial_\mu \psi \partial_\nu \xi-g_{\mu\nu}\partial_\alpha\psi\partial^\alpha\xi\bigg)\,,
\end{align}
which can be straightforward calculated in the homogeneous background as follows
\begin{equation} \label{cross}
I_{\mu\nu}(t)=\left\{ \begin{aligned}
          & -\frac{1}{2\sqrt{-g}}\int^{t}_{+\infty}dt'\sqrt{-g(t')}R(t')\int_{-\infty}^{t}dt''\sqrt{-g(t'')}R(t'')\,, \qquad \mu=\nu=0 \\
          & \frac{1}{2}\frac{g_{\mu\nu}}{\sqrt{-g}}\int^{t}_{+\infty}dt'\sqrt{-g(t')}R(t')\int_{-\infty}^{t}dt''\sqrt{-g(t'')}R(t'')\,, \qquad \mathrm{other~ cases}
                          \end{aligned} \right.
                          \end{equation}
It is clear that Eq.~(\ref{cross}) exactly coincides with the second line in the original equation of motion~(\ref{eomorigin}). Hence, on the background level, once the auxiliary fields are identified with the terms in the original nonlocal equations and we remove the spurious degrees of freedom, the local field equations are equivalent to the original nonlocal forms.

In a similar way with the discussion of $\psi$, a solution for $\xi$ from the localized equation (\ref{st3}) involves the homogeneous solution which is absent in the original non-local theory. Hence, one should directly work with Eq.~(\ref{xipower}): for power-law solutions, it can be integrated as
\begin{align}\label{xipowersol}
\xi(t)=\frac{6s(2s-1)}{3s-1}\left\{\frac{1}{1-3s}\left[\left(\frac{t}{t_A}\right)^{1-3s}-1\right]-\ln\left(\frac{t}{t_A}\right)\right\}\,,
\end{align}
where $t_A$ is the boundary for integration of the advanced Green's function. There appears an interesting observation that mathematically, one should put $t_A\rightarrow+\infty$ which causes a divergence in Eq.~(\ref{xipowersol}). A possible way to avoid this is choose $s\rightarrow1/2$ or $s\rightarrow0$ so that the solution for $\xi$ converges. This necessarily means that the Universe will go back to the radiation-dominated epoch, or even Minkowskian in the future. We can understand this result intuitively: in both stages, the Ricci scalar vanishes so that the nonlocal modification disappears, hence the divergence problem could be avoided in these two cases.

Using the identification $\psi=\Box^{-1}_RR$ and $\xi=\Box^{-1}_AR$, the localised field equation (\ref{st1}) is equivalent to the non-local theory without introducing any spurious degree of freedom. In the FLRW background, Eq.~(\ref{st1}) reduces to the Hamiltonian and momentum constraints, respectively:
\begin{align}
\label{einstein1} 0 &=~ - 3 H^2\left(1 + \psi + \xi\right) -
\frac{1}{2}\dot\xi \dot\psi
 - 3H\left(\dot\psi + \dot\xi\right)\, ,\\
\label{einstein2} 0 &=~ \left(2\dot H + 3H^2\right) \left(1 +
\psi + \xi\right) - \frac{1}{2}\dot\xi \dot\psi +
\left(\frac{d^2}{dt^2} + 2H \frac{d}{dt} \right) \left( \psi +\xi
\right)\,.
\end{align}
It can be observed from the above equations that when the Ricci scalar vanishes $R=0$, the only vacuum solution is the Minkowsky spacetime where the index $s=0$.\footnote{When $s=1/2$, from Eq.~(\ref{Adxi}) we have $\psi=\xi=0$. Inserting this into Eq.~(\ref{einstein1}) with $H=1/(2t)$, we find that the equation does not hold.} Hence, for vacuum power-law solutions, as a consequence of the acausality problem, the divergence problem discussed below Eq.~(\ref{xipowersol}) cannot be avoided unless the Universe approaches the Minkowsky spacetime in the infinite future.

\section{General case}\label{general}
Based on the discussion in the previous section, now we consider a general case where an arbitrary function of $\Box^{-1}R$ arises in the action
\begin{align}\label{generalf}
S_{f}=\int d^4x\sqrt{-g(x)}R(x)f\left(\mathrm{X}, \mathrm{Y}\right)\,,
\end{align}
where $\mathrm{X}\equiv\left(\Box^{-1}_RR\right)[x]$, $\mathrm{Y}\equiv\left(\Box^{-1}_AR\right)[x]$ and $f(\mathrm{X}, \mathrm{Y})$ is an arbitrary function of its variables $\mathrm{X}$ and $\mathrm{Y}$. In a similar way with the analysis in Section.~\ref{EOMdeduce}, in the homogeneous background, the variation of Eq.~(\ref{generalf}) with respect to $g_{\mu\nu}$ can be obtained as follows
\begin{align}\label{eomf}
\frac{\delta S_{f}}{\delta g^{\mu\nu}}&=\sqrt{-g}\left(R_{\mu\nu}-\frac{1}{2}g_{\mu\nu}R
-\nabla_{\mu}\nabla_{\nu}+g_{\mu\nu}\Box\right)\left[f+\Box^{-1}_A(Rf_{\mathrm{X}})+\Box^{-1}_R(Rf_{\mathrm{Y}})\right]\nonumber\\
&-\frac{g_{\mu\nu}}{2\sqrt{-g}}\int_t^{+\infty}dt'\sqrt{-g(t')}R(t')f_{\mathrm{X}}(t')\int_{-\infty}^{t}dt''\sqrt{-g(t'')}R(t'')\nonumber\\
&-\frac{g_{\mu\nu}}{2\sqrt{-g}}\int^t_{-\infty}dt'\sqrt{-g(t')}R(t')f_\mathrm{Y}(t')\int^{+\infty}_{t}dt''\sqrt{-g(t'')}R(t'')\,,
\end{align}
where a subscript index means the partial derivative with respect to it: $f_\mathrm{X}\equiv\partial f/\partial\mathrm{X}$. It is clear that because of the symmetrisation of the retarded and advanced Green's function in the EOM, the advanced Green's function cannot be eliminated from any construction of function $f(\mathrm{X}, \mathrm{Y})$. This means that future information is always needed to find the evolution of the current universe, hence this implies the acusality problem in nonlocal gravity theories in general.

\section{Conclusion}
We considered the acausality problem in the nonlocal gravity theory. Starting from the action for a nonlocal gravity theory, the equations of motion (EOM) was derived by the variation principle. It was found that the advanced Green's function would always appear in the EOM, regardless the definition of the nonlocal operator, or the construction of the nonlocal functions in the action. This means that cosmology described by the nonlocal gravity theory is determined not only by the past history, but also by the future evolution of the Universe.

On the other hand, by introducing extra scalars, the nonlocal gravity theory could be written in the localised form in the biscalar-tensor presentation. We made a comparison of the original EOM to its biscalar-tensor presentation and identify the extra scalars with the nonlocal terms in the original EOM. We found that one of the extra scalars would be identified with the term associated with the advanced Green's function. Generally speaking, the solutions given in the biscalar-tensor presentation contain homogeneous solutions which would lead to the ``ghost''-like instability problem~\cite{Nojiri:2010pw,ZS:2011,BNOS:2011,deS:2015}. Hence, in the nonlocal gravity theory, one would either result in the acausality problem in the original frame, or the instability problem in the biscalar-tensor presentation.

The acausality problem may imply that the nonlocal gravity theory defined in Eq.~(\ref{actiong}) (or generally defined in Eq.~(\ref{generalf})) is not a fundamental theory. An analog example was discussed in Refs.~\cite{MM:2014,FMM:2014gh}, where the dynamics of in-out matrix element of quantum fields was provided by the path integral approach. The corresponding classical equations contain both the retarded and advanced Green's functions, but this does not necessarily imply disaster since the in-out matrix elements do not directly correspond to the measurable physical quantities. Similarly, one may expect that in a semiclassical approach to gravity, the in-out matrix $\langle0_{\rm in}|g_{\mu\nu}|0_{\rm out}\rangle$ should not be interpreted as an effective metric. One approach to this notion is that in the effective equations the $\Box^{-1}$ operator should be replaced by the retarded Green's function only~\cite{SW:2003}, which may correspond to the in-in matrix element $\langle0_{\rm in}|g_{\mu\nu}|0_{\rm in}\rangle$~\cite{M:2013}.\footnote{It should be noted that except for some simple cases, this replacement does not exactly correspond to the in-in
formulism~\cite{Deser:2007jk}. For more detailed discussion, see ref.~\cite{TW:2014}.}

On the other hand, one possible way to obtain causal nonlocal equations is
to make a separation between long-wavelength and short-wavelength modes and smooth the equations of motion to extract the effective evolution equation for the long-wavelength mode~\cite{CLP:2014}. It would be very interesting to explore this possibility or to find an underlying fundamental theory that gives causal nonlocal equations motion in the future work.

\section*{Acknowledgment}
We would like to thank Takahiro Tanaka for valuable discussion. KK thanks M. Maggiore for bringing Ref.~\cite{M:2013} to our attention on the discussion about the acausality problem. This work is supported in part by MEXT KAKENHI Grant Number 15H05888. YZ and GZ are supported by the Strategic Priority Research Program
``The Emergence of Cosmological Structures" of the Chinese
Academy of Sciences, Grant No. XDB09000000.
KK is supported by the UK Science and Technology Facilities Council grants ST/K00090/1 and the European Research Council grant through 646702 (CosTesGrav).

%%%%%%%%%%%%%%%
\section*{Appendix: Appearance of the advanced Green's function in the FLRW metric}\label{scalarapp}
%%%%%%%%%%%%%%%
%
%Before doing variation, I should note that by which Green's function
%the nonlocal operator is defined. To preserve the causality, I
%define the nonlocal operator by using retarded Green's function
%$G_R(x, x')$:
%\begin{align}
%\left(\Box^{-1}_Rf\right)[x]\equiv\int d^4x'\sqrt{-g(x')}f(x')G_R(x,
%x')\,,
%\end{align}
%while it should be noted that under the change of orders of
%variables for integration $x\leftrightarrow x'$, the retarded
%Green's function becomes an advanced one $G_A(x', x)$. Now we
%consider the term
%\begin{align}
%&~~~\int d^4x\sqrt{-g(x)}F[g(x)]\left(\Box^{-1}_R\frac{\delta
%R}{\delta
%g^{\mu\nu}(y)}\right)[x]\nonumber\\
%&=\int\int d^4x~d^4x'\sqrt{-g(x)}\sqrt{-g(x')}F[g(x)]G_R(x,
%x')\frac{\delta R(x')}{\delta g^{\mu\nu}(y)}\nonumber\\
%&=\int\int d^4x'~d^4x\sqrt{-g(x')}\sqrt{-g(x)}F[g(x')]G_R(x',
%x)\frac{\delta R(x)}{\delta g^{\mu\nu}(y)}\nonumber\\
%&=\int d^4x\sqrt{-g(x)}\frac{\delta R(x)}{\delta g^{\mu\nu}(y)}\int
%d^4x'\sqrt{-g(x')}F[g(x')]G_R(x', x)\nonumber\\
%&=\int d^4x\sqrt{-g(x)}\frac{\delta R(x)}{\delta g^{\mu\nu}(y)}\int
%d^4x'\sqrt{-g(x')}F[g(x')]G_A(x, x')\nonumber\\
%&=\int d^4x\sqrt{-g(x)}\frac{\delta R(x)}{\delta
%g^{\mu\nu}(y)}\left(\Box^{-1}_AF[g]\right)[x]\,, \label{tildeg}
%\end{align}
%which implies that the `partial integration' will make the nonlocal
%operator become an advanced one, thus causes the acausal problem.

In Sec.~\ref{sec2}, using the property that, under the replacement of variables $x\leftrightarrow x'$, the retarded Green's function changes to the advanced one $G_R(x, x')\leftrightarrow G_A(x', x)$, the appearance of the advanced Green's in the EOM was shown explicitly. In this appendix, without using this property, we will show the same result in the homogeneous background described by the FLRW metric. In this case, using Eq.~(\ref{boxg}), the nonlocal operator $\Box^{-1}$ can
be expressed as follows,
\begin{equation}
\label{boxFLRW}
\left(\Box^{-1}\phi\right)[t]=-\int_{t_1}^t\frac{dt'}{a^3(t')}\int_{t_0}^{t'}dt''a^3(t'')\phi(t'')\,,
\end{equation}
where $t_0$ and $t_1$ are two boundaries for integration. Hence, the variation of the action (\ref{scalaraction}) can be calculated as
\begin{align}
\frac{\delta S_\phi}{\delta\phi(\tilde{t})}
&=-\frac{\delta}{\delta\phi(\tilde{t})}\left\{\int_{-\infty}^{+\infty}dt~
a^3(t)\phi(t)\left[\int_{t_1}^t\frac{dt'}{a^3(t')}\int_{t_0}^{t'}dt''a^3(t'')\phi(t'')\right]\right\}\label{part0}\\
&=-\int_{-\infty}^{+\infty}dt~
a^3(t)\delta(t-\tilde{t}~)\int_{t_1}^t\frac{dt'}{a^3(t')}\int_{t_0}^{t'}dt''a^3(t'')\phi(t'')\label{part1}\\
&~~~-\int_{-\infty}^{+\infty}dt~
a^3(t)\phi(t)\int_{t_1}^t\frac{dt'}{a^3(t')}\int_{t_0}^{t'}dt''a^3(t'')\delta(t''-\tilde{t}~)\,.\label{part2}
\end{align}

For definiteness, we set $t_0\rightarrow-\infty$ and
$t_1\rightarrow-\infty$ so that at the level of action, the
$\Box^{-1}$ is defined in terms of the retarded Green's function. We first consider the term (\ref{part2}). Using the expression
\begin{eqnarray}\label{step1}
\int_{-\infty}^t\frac{dt'}{a^3(t')}\int_{-\infty}^{t'}dt''a^3(t'')\delta(t''-\tilde{t}~)
=a^3(\tilde{t}~)\theta\left(t-\tilde{t}\right)\int_{\tilde{t}}^t\frac{dt'}{a^3(t')}\,,
\end{eqnarray}
where the step function $\theta\left(t-\tilde{t}\right)$ is defined
as
\begin{equation} \label{eq:1}
\theta\left(t-\tilde{t}\right)=\left\{ \begin{aligned}
          &~1,\qquad t\geq\tilde{t}\,, \\
                  &~0, \qquad t<\tilde{t}\,,
                          \end{aligned} \right.
                          \end{equation}
Eq.~(\ref{part2}) can be evaluated as follows:
\begin{align}\label{interstep}
&~~~-\int_{-\infty}^{+\infty}dt~
a^3(t)\phi(t)\int_{-\infty}^t\frac{dt'}{a^3(t')}\int_{-\infty}^{t'}dt''a^3(t'')\delta(t''-\tilde{t}~)\nonumber\\
&=-a^3(\tilde{t}~)\int_{\tilde{t}}^{+\infty}dt~\partial_t\left[\int_{T}^t
dt''~a^3(t'')\phi(t'')\right]\int_{\tilde{t}}^t\frac{dt'}{a^3(t')}\nonumber\\
&=-a^3(\tilde{t}~)\left\{\int_{\tilde{t}}^{+\infty}dt~\partial_t\left[\int_{T}^t
dt''~a^3(t'')\phi(t'')\int_{\tilde{t}}^t\frac{dt'}{a^3(t')}\right]-\int_{\tilde{t}}^{+\infty}dt\int_{T}^t
dt''~a^3(t'')\phi(t'')\partial_t\left[\int_{\tilde{t}}^t\frac{dt'}{a^3(t')}\right]\right\}\nonumber\\
&=-a^3(\tilde{t}~)\left\{\left[\int_{T}^t
dt''~a^3(t'')\phi(t'')\int_{\tilde{t}}^t\frac{dt'}{a^3(t')}\right]_{t=\tilde{t}}^{t=+\infty}-\int_{\tilde{t}}^{+\infty}\frac{dt}{a^3(t)}
\int_{T}^t dt''~a^3(t'')\phi(t'')\right\}\,,
\end{align}
where it should be noted that in the second line, we inserted
\begin{eqnarray}
a^3(t)\phi(t)=\partial_t\left[\int_{T}^t
dt''~a^3(t'')\phi(t'')\right]\,,\label{stepin1}
\end{eqnarray}
with $T$ a constant for integral boundary, and we integrate by parts
in the third step. For simplicity, we set $T=\tilde{t}$ so that
Eq.~(\ref{interstep}) becomes
\begin{align}\label{stepin2}
(\ref{interstep})&=-a^3(\tilde{t}~)\left\{\int_{\tilde{t}}^{+\infty}
dt''~a^3(t'')\phi(t'')\int_{\tilde{t}}^{+\infty}\frac{dt'}{a^3(t')}-\int_{\tilde{t}}^{+\infty}\frac{dt}{a^3(t)}
\int_{\tilde{t}}^t dt''~a^3(t'')\phi(t'')\right\}\nonumber\\
&=-a^3(\tilde{t}~)\int_{\tilde{t}}^{+\infty}\frac{dt}{a^3(t)} \int_{t}^{+\infty}
dt''~a^3(t'')\phi(t'')\nonumber\\
&=a^3(\tilde{t}~)\left(\Box^{-1}_A\phi\right)[\tilde{t}]\,.
\end{align}

On the other hand, under $t_0\rightarrow-\infty$ and
$t_1\rightarrow-\infty$, Eq.~(\ref{part1}) becomes:
\begin{align}\label{part3}
&~~~~-\int_{-\infty}^{+\infty}dt~
a^3(t)\delta(t-\tilde{t}~)\int_{-\infty}^t\frac{dt'}{a^3(t')}\int_{-\infty}^{t'}dt''a^3(t'')\phi(t'')\nonumber\\
&=
-a^3(\tilde{t})\int_{-\infty}^{\tilde{t}}\frac{dt'}{a^3(t')}\int_{-\infty}^{t'}dt''a^3(t'')\phi(t'')\nonumber\\
&=a^3(\tilde{t}~)\left(\Box^{-1}_R\phi\right)[\tilde{t}]\,.
\end{align}

Inserting (\ref{stepin2}) and (\ref{part3}) into
(\ref{part0}), one obtains that
\begin{align}\label{final}
\frac{\delta S_\phi}{\delta\phi(\tilde{t})}=a^3(\tilde{t}~)\bigg\{\left(\Box^{-1}_A\phi\right)+\left(\Box^{-1}_R\phi\right)\bigg\}[\tilde{t}]\,.
\end{align}
where the nonlocal operator defined by the advanced Green's function
is notated as $\Box^{-1}_A$. The same result can be obtained when the nonlocal operator in action (\ref{scalaraction}) is defined by the advanced Green's function.

Thus, for a nonlocal action where the nonlocal operator acts on a scalar field such as Eq.~(\ref{scalaraction}), we have shown that in the homogeneous background described by the FLRW metric, no matter how the nonlocal operator is defined, the corresponding equation of motion always contains the
advanced Green's function, which breaks the causality.

\end{document}